\documentclass[a4paper]{jpconf}
\usepackage{graphicx}
\usepackage{mathrsfs}
\usepackage{dsfont}
\usepackage{xcolor}
\usepackage{amssymb}
\usepackage[T1]{fontenc}
\usepackage[font=small,labelfont=bf,tableposition=top]{caption}
\DeclareMathAlphabet{\mathpzc}{OT1}{pzc}{m}{it} %define calligraphic font for small
\begin{document}

\title{Constructing realistic alpha cluster channels}

\author{K. Kravvaris and A. Volya}

\address{Florida State University, Tallahassee, FL 32306, USA}

%\ead{kk11f@my.fsu.edu}

\begin{abstract}
We present techniques that allow for $\alpha$-cluster channels with realistic $\alpha$-particle wave functions from No Core Shell Model calculations to be constructed. We compare results of several clustering calculations with realistic $\alpha$ wave functions to those assuming a trivial
$(0s)^4$ structure. 
\end{abstract}

In this work we report our progress in addressing specific questions related to studies of $\alpha$ clustering from the perspective of the nuclear shell model approach; 
our discussion follows a series of works in Refs. [1-7],
%\cite{Mang:1957,Chung:1978,Smirnov:1977,Chung:1978,Grigorescu:1993, Tchuvilsky:1995,Navratil:2004,Nemetz:1988}, 
and \cite{Volya:2015}, in particular. 
Let us start by reviewing several key elements of the shell model approach to clustering. 
The use of harmonic oscillator (HO) basis, which we adopt for this work, allows for a formal separation of the center-of-mass (CM) degrees of freedom. Having an additional HO confining CM potential allows for factorization of the CM degree of freedom, and the full 
$A$-nucleon wave function appears as a product 
\begin{equation}
\Psi=\phi_{{n}\ell\mathpzc{m}}({\bf R})\, \Psi' 
\label{1}
\end{equation}
of the 
HO wave function $\phi_{{n}\ell\mathpzc{m}}({\bf R})$ that depends only on the CM variables ${\bf R}$ 
and the translationally invariant wave function $\Psi' $, which is a function of relative coordinates only. 
In our notation $n$ is the number of oscillator quanta. 
Thus, in eq. (\ref{1}) the total number of oscillator 
excitation quanta is shared between the CM and intrinsic degrees of freedom 
$N=n+N'.$

In the traditional approach to clustering the solutions for parent and daughter systems are both obtained in the form (\ref{1}) where $n=0.$ 
It is then assumed than an intrinsic state of the $\alpha$ particle is $(0s)^4$ configuration and thus $N'_{\alpha}=0.$  This approximation simplifies the approach significantly because then the wave function of the $\alpha$ particle in a given HO CM state, the channel, can be written as 
\begin{equation}
\Psi_\alpha=\phi_{{n}\ell\mathpzc{m}}({\bf R})\, \Psi_\alpha'= \sum_\eta X^\eta_{{n}\ell}
\,\Phi^\eta_{(n,0):\ell \mathpzc{m}}.
\label{2}
\end{equation}
The expansion goes over all possible $A=4$ nucleon configurations (often also referred to as partitions), labeled by $\eta$ which have a stretched SU(3) symmetry $(n,0),$ full permutational symmetry, and spin and isospin quantum numbers $S=T=0.$ The expansion cluster coefficients (CC) $X^\eta_{{n}\ell}$ are known analytically \cite{Smirnov:1977, Volya:2015}
%\cite{Ichimura:1973,Smirnov:1977,Smirnov:1983},
\begin{equation}
X^\eta_{{n}\ell} =\sqrt{\frac{1}{4^n}\,\frac{n!}{\prod_i ({ n}_i !)^{\alpha_i}}\,
\frac{4!}{\prod_i \alpha_i !} }. \label{eq:CC1}
\end{equation}
Here  $\alpha_i$ refers to the number of particles on an oscillator shell  $n_i.$ 
Then the fractional parentage coefficient between parent and daughter states ${\cal F}^\eta_{nL}\equiv \langle \Psi_P |   \Phi^{\eta\dagger}_{(n,0):L} | \Psi_D   \rangle  $ can be evaluated with the standard shell model techniques utilizing  operator formalism of the second quantization. 
The translationally invariant fractional parentage coefficient requires an additional recoil coefficient 
that emerges as an oscillator bracket from 
recouping of the center of mass variables of the daughter system and $\alpha$ particle into a relative coordinate and a common CM that coincides with the parent system. The whole procedure and additional renormalization strategy is discussed in Ref. \cite{Volya:2015}.

The purpose of this work is to present an approach that does not require a simple $(0s)^4$ structure for the $\alpha$ particle. We obtain the $\alpha$ wave function, restricted by the maximum total number of HO quanta $N_{\rm max}$ using the No Core Shell Model (NCSM) scheme, we use interactions from \cite{Shirokov:2008}.
The realistic $\alpha$ wave function invalidates the expansion in  eq. (\ref{2}), and renders the analytic result in (\ref{eq:CC1}) useless. Nevertheless,  the proposed new technique is just as effective numerically since it does not require construction of SU(3) operators 
$\Phi^\eta_{(n,0):\ell \mathpzc{m}}.$

We start with an $\alpha$ particle wave function being calculated using the NCSM approach.  
Upon application of the Glockner-Lawson procedure this produces a state of the same type as in eq. (\ref{1}), where realistic $\alpha$ particle is in the CM HO state with $n=0.$ Then using sequential applications of the the CM creation and annihilation operators we obtain states where the $\alpha$ particle is in the CM state (\ref{1}) with any desired CM HO quantum numbers. 
The CM creation operators (and annihilation correspondingly) are 
defined in the usual way as 
\begin{equation}
\mathcal{B}_m^\dagger = 
\frac{1}{\sqrt{2AM\Omega\hbar}}  (A M \Omega R_m- iP_m)=\frac{1}{\sqrt{A}}\displaystyle\sum_{a=1}^{A} b^\dagger_{a\,m}
\end{equation}
where $m$ denotes a specific magnetic projection of vectors 
and $b^\dagger_{am}$ raises the quanta of the $a$-th particle. 
The operator is easily constructed using an isoscalar mass-density dipole E1 operator 
\begin{equation}
D_m=\sqrt{\frac{4\pi}{3}} \sqrt{\frac{\hbar}{2AM\Omega}}(\mathcal{B}_m^\dagger + \mathcal{B}_m)
\end{equation}
and by taking the part that increases the number of quanta. 

In order to boost the CM wave function to a particular state with $n=2p+\ell$ quanta, where $p$ is the number of nodes and  $\ell$ the angular momentum, one can apply various combinations of 
creation operators   $\mathcal{B}_m^\dagger.$ The number of nodes can be increased, keeping rotational quantum numbers unchanged using a scalar combination of two creation operators 
\begin{equation}
\mathcal{B}^\dagger \cdot \mathcal{B}^\dagger \equiv
 \left (\mathcal{B}_{+1}^\dagger \mathcal{B}_{-1}^\dagger + 
\mathcal{B}_{-1}^\dagger \mathcal{B}_{+1}^\dagger - 
\mathcal{B}_{0}^\dagger \mathcal{B}_{0}^\dagger \right ),
\end{equation}
 \begin{equation}
\mathcal{B}^\dagger \cdot \mathcal{B}^\dagger \phi_{{n}\ell\mathpzc{m}}({\bf R})\,  = \sqrt{{(n-\ell+2)(n+\ell+3)}}/4\,\,\phi_{{n+2}\ell\mathpzc{m}}({\bf R}).
\end{equation}
In order to increase the angular momentum  $\ell$ while keeping the number of nodes in the wave function fixed one can act with  $\mathcal{B}_m^\dagger.$ The simplest strategy is to build an aligned state where $m=\ell$
\begin{equation}
\mathcal{B}_{+1}^\dagger \phi_{{n}\ell\ell}({\bf R}) =  \sqrt{\frac{(\ell+1)(n+\ell+3)}{4(2\ell+3)}}  \phi_{{n+1}\ell+1 \ell+1}({\bf R}).
\end{equation}
The CM angular momentum operator is a vector construction from CM quanta 
creation and annihilation operators $\mathcal{B}^\dagger \times \mathcal{B}.$ In particular, the usual lowering operator required to obtain the desired magnetic projection is 
\begin{equation}
\mathcal{L}_{-} \phi_{{n}\ell\mathpzc{m}}({\bf R}) =
4\sqrt{2}\left ( \mathcal{B}^\dagger_{-1} \mathcal{B}_{0}-\mathcal{B}^\dagger_{0} \mathcal{B}_{-1} \right ) \phi_{{n}\ell\mathpzc{m}}({\bf R}) = \sqrt{(l+m)(l-m+1)}\phi_{{n}\ell\mathpzc{m}-1}({\bf R}).
\end{equation}

\begin{minipage}[h]{\textwidth}
	\begin{minipage}[t]{0.40\textwidth}
		\begin{tabular}{l c c}
			\br
			Configuration & $N_{\rm max}=0$ & $N_{\rm max}$ = 4  \\ 
			\mr
			$(sd)^4$    		& 0.038 & 0.035\\[0.3em]
			$(p)(sd)^2(pf)$  		& 0.308 & 0.282\\[0.3em]
			$(p)^2(pf)^2$ 		& 0.103 & 0.094\\[0.3em]
			$(p)^2(sd)(sdg)$		& 0.154 & 0.141\\[0.3em]
			$(s)^2(sd)(sdgi)$	& 0.000 & 0.005\\[0.3em]
			$(p)(sd)(pf)(sdg)$	& 0.000 & 0.009\\[0.3em]
			\br
		\end{tabular}
		\label{tab:cc}
		\captionof{table}{Select configuration content of NCSM wave functions for $^4$He with $\hbar\Omega = 20$ MeV boosted by 8 quanta ($L=0$).}
	\end{minipage}
	\qquad
	\quad
	\begin{minipage}[t]{0.44\textwidth}
		\begin{tabular}{ c c c c }
			\br
			A$_{P}\rightarrow$A$_D$ & $N_{\rm max}=0$  & $N_{\rm max}=6$ & Exp.\\
			\mr
			$^{20}$Ne$\rightarrow^{16}$O   	&0.755  &0.827& 1\\[0.3em]
			$^{22}$Ne$\rightarrow^{18}$O   	&0.481  &0.563& 0.37\\[0.3em]
			$^{24}$Mg$\rightarrow^{20}$Ne 	&0.411  &0.519& 0.66\\[0.3em]
			$^{26}$Mg$\rightarrow^{22}$Ne 	&0.439  &0.548& 0.20\\[0.3em]
			$^{28}$Si$\rightarrow^{24}$Mg  	&0.526 &0.575& 0.33\\[0.3em]
			$^{30}$Si$\rightarrow^{26}$Mg  	&0.555 &0.600& 0.55\\[0.3em]
			\br
		\end{tabular}	
      		\captionof{table}{Spectroscopic Factors for sd shell ground states using SU(3) and realistic $\alpha$ cluster
      		 wave functions ($\hbar\Omega = 14$ MeV).}
      		\label{tab:sf}
	\end{minipage}
\end{minipage}

In Table 1 we compare the weights of select configuration components of $\alpha$ wave functions 
from NCSM calculations with $N_{\rm max}=4$
that have been 
CM-boosted for $n=8$ using the raising operators with the $N_{\rm max}=0$ case. The  
$N_{\rm max}=0$ case corresponds to  
$(0s)^4$ $\alpha$ configuration and reproduces the squared CC in eq. (\ref{eq:CC1}). 
For configurations with ${N=\sum_i n_i \alpha_i =8}$ the ratio between the two columns is proportional to 
the weight of the $(0s)^4$ configuration in the $\alpha$ wave function.  This weight depends on the 
oscillator frequency $\Omega;$ here $\hbar\Omega=20$MeV. 
The last two lines in Table 1, show $N=10$ configurations that are only present in a realistic $\alpha$  wave function that is different from $(0s)^4.$

In Table 2 we demonstrate the effect of using a realistic $\alpha$ cluster in spectroscopic calculations. Here, 
$\hbar\Omega=14$ MeV that is a more typical value for the  traditional shell model. 
The spectroscopic factors are calculated using the procedure and $sd$ shell model Hamiltonian 
outlined in \cite{Volya:2015}. 
For $N_{\rm max}=0$ in the $sd$ shell the only SU(3) component that contributes is the (8,0) irreducible representation; 
this is not the case  in a more realistic situation with $N_{\rm max}=6.$ 
Here, channels with $p=4,3,2,1$ all contribute to  the $(sd)^4$ valence configuration. 

In summary, in this work we put forward a new strategy for clustering calculations using the
oscillator-based shell model approach. The strategy is efficient numerically, does not rely on algebraic techniques, and allows to treat realistic wave functions of clusters from NCSM. Select simple examples presented here demonstrate the procedure, show the effects of more complex $\alpha$ wave functions and highlight the limit when our approach reduces to the algebraic method used previously.   
\ack
We thank Yu. M. Tchuvil'sky for collaboration and helpful comments.
This material is based upon work supported by the U.S. Department of Energy Office of Science, Office of Nuclear Physics under Grant No. DE-SC0009883

\section*{References}
%\bibliographystyle{iopart-num}
%\bibliography{su3_prc}

\end{document}